\documentclass[aps,pre,reprint,twocolumn,superscriptaddress,amsmath,showpacs]{revtex4-1}
\pdfoutput = 1

\usepackage{color,layouts,graphicx,setspace,tabularx,bm,diagbox}
\definecolor{darkblue}{rgb}{0,0,0.6}
\definecolor{darkred}{rgb}{0.6,0,0}
\definecolor{darkgreen}{rgb}{0,0.6,0}

\usepackage[colorlinks=true,urlcolor=darkblue,citecolor=darkblue,linkcolor=darkred,hyperfootnotes=false]{hyperref}
\usepackage[normalem]{ulem}

\newcommand{\stkout}[1]{\ifmmode\text{\sout{\ensuremath{#1}}}\else\sout{#1}\fi}
\newcolumntype{C}{>{\centering\arraybackslash}X}

\newcommand{\stirling}{\genfrac{[}{]}{0pt}{}}

\begin{document}
\title{Role of hubs in the synergistic spread of behavior}


\author{Yongjoo Baek}
\email[These authors contributed equally to this work.]{}
\affiliation{DAMTP, Centre for Mathematical Sciences, University of Cambridge, Cambridge CB3 0WA, United Kingdom}

\author{Kihong Chung}
\email[These authors contributed equally to this work.]{}
\affiliation{Natural Science Research Institute, Korea Advanced Institute of Science and Technology, Daejeon 34141, Korea}

\author{Meesoon Ha}
\email[Corresponding author: ]{msha@chosun.ac.kr}
\affiliation{Department of Physics Education, Chosun University,
Gwangju 61452, Korea}

\author{Hawoong Jeong}
\affiliation{Department of Physics and Institute for the
BioCentury, Korea Advanced Institute of Science and Technology,
Daejeon 34141, Korea}

\author{Daniel Kim}
\affiliation{Natural Science Research Institute, Korea Advanced Institute of Science and Technology, Daejeon 34141, Korea}

\date{\today}

\begin{abstract}
The spread of behavior in a society has two major features: the synergy of multiple spreaders and the dominance of hubs. While strong synergy is known to induce mixed-order transitions (MOTs) at percolation, the effects of hubs on the phenomena are yet to be clarified. By analytically solving the generalized epidemic process on random scale-free networks with the power-law degree distribution $p_k \sim k^{-\alpha}$, we clarify how the dominance of hubs in social networks affects the conditions for MOTs. Our results show that, for $\alpha < 4$, an abundance of hubs drive MOTs, even if a synergistic spreading event requires an arbitrarily large number of adjacent spreaders. In particular, for $2 < \alpha < 3$, we find that a global cascade is possible even when only synergistic spreading events are allowed. These transition properties are substantially different from those of cooperative contagions, which are another class of synergistic cascading processes exhibiting MOTs.
\end{abstract}



\maketitle


{\em Introduction.} There has been a growing body of literature on {\em mixed-order transitions} (MOTs), which qualify as both continuous and discontinuous phase transitions depending on the chosen order parameter. Such transitions appear in many different contexts, such as DNA unzipping~\cite{PolandJCP1966,CausoPRE2000,KafriPRL2000}, Ising spins with long-range interactions~\cite{BarPRL2014,*BarJSM2014}, and various percolation models with biased merger of clusters~\cite{AraujoEPJST2014}. A common aspect of these systems is the existence of long-range interactions which encourage global ordering over a finite fraction of the system at criticality~\cite{BarPRL2014,*BarJSM2014}.

Recently added to the list are various models of cascades with synergistic spreading rules involving cooperation between different contagions~\cite{GrassbergerPRE2016,LChenEPL2013,WCaiNP2015,CuiPRE2017}, weakened individuals~\cite{KChungPRE2016,DLeeSciRep2017,WChoiPRE2017a,*WChoiPRE2017b,*WChoiPRE2018,BizhaniPRE2012,GrassbergerPRL2018,JanssenPRE2004,*JanssenEPL2016,*JanssenJPA2017,KChungPRE2014}, or multiple spreading thresholds~\cite{BMinSciRep2018}. If each transmission occurs independently without synergy, the cascade exhibits a continuous percolation transition~\cite{PastorSatorrasRMP2015}. In contrast, with sufficiently strong synergy, the transition can be a MOT: a continuous transition of the probability of a global cascade coincides with a discontinuous jump of the cascade size. Moreover, the lines of MOTs and purely continuous transitions join at a tricritical point (TCP) with its own critical properties~\footnote{Rigorously speaking, a TCP is an endpoint of the coexistence line shared by three different phases. It is unclear whether the same is true for cooperative contagions, but we follow the casual definition of a TCP as a continuous transition point at the intersection between continuous and discontinuous transition lines.}. Again, the long loops of the substrate, through which different spreading pathways cross each other, facilitate global cascades at the MOTs~\cite{WCaiNP2015,DLeeSciRep2017}.

A natural question arises on how the conditions for MOTs depend on the structure of the underlying substrate. In homogeneous structures, such as lattices~\cite{GrassbergerPRL2018,BizhaniPRE2012,GrassbergerPRE2016,JanssenPRE2004,*JanssenEPL2016,*JanssenJPA2017,LChenEPL2013}, Poissonian random networks~\cite{BizhaniPRE2012,KChungPRE2016,WCaiNP2015,GrassbergerPRE2016,DLeeSciRep2017,WChoiPRE2017a,*WChoiPRE2017b,*WChoiPRE2018,LChenEPL2013,BMinSciRep2018}, and modular networks~\cite{KChungPRE2014}, a MOT requires sufficiently strong synergy between two spreaders and dimension greater than two~\cite{BizhaniPRE2012,GrassbergerPRL2018}. However, cascades typically occur on heterogeneous structures: for instance, social networks feature a significant fraction of highly-connected individuals called {\em hubs}, whose existence is typically modeled by {\em scale-free networks} (SFNs) with a power-law distribution $p_k \sim k^{-\alpha}$ (with $\alpha > 2$) of the number of neighbors $k$ (called {\em degree})~\cite{Newman2010}. Since SFNs with a greater variance of $k$ contain more loops~\cite{BianconiJSM2005}, $\alpha$ can be a major determinant of the conditions for MOTs. For cooperative contagions on SFNs, a heterogeneous mean-field approach~\cite{CuiPRE2017} showed that a discontinuous jump of the cascade size is possible for $\alpha > 3$ given sufficiently strong synergy, but not for $2 < \alpha < 3$; however, whether the same statement holds for general kinds of synergy remains to be clarified.

In this study, we show that the synergistic spread of behavior exhibits substantially different transition phenomena for small values of $\alpha$. As empirically observed~\cite{CentolaSci2010}, social reinforcement induces a large boost in the spread of a behavior if the target individual has sufficiently many adjacent spreaders. As a simple model incorporating this feature, we study the generalized epidemic process (GEP) with the {\em synergy threshold} $n \ge 2$, in which the spreading probability changes when the number of spreading neighbors is greater than or equal to $n$, extending the original version limited to $n = 2$~\cite{BizhaniPRE2012}. In the sense that the cluster is formed by a mixture of single-node and multi-node mechanisms, our model can be considered a cascading-process analog of the heterogeneous $k$-core percolation~\cite{BaxterPRE2011}, which is a pruning process. We analytically show that, for $2 < \alpha < 4$, an abundance of hubs enable MOTs for arbitrarily large $n \ge 2$. In contrast to cooperative contagions, the cascade size exhibits a discontinuous jump even for $2 < \alpha < 3$ in a manner similar to the abrupt appearance of a giant heterogeneous $k$ core with $k \ge 3$ on the same SFNs~\cite{BaxterPRE2011}. While the near-TCP scaling exponents for $\alpha > 3$ remain identical to those of cooperative contagions~\cite{CuiPRE2017}, a new set of exponents can be identified for $2 < \alpha < 3$. 

\begin{figure*}[]
\includegraphics[width=\textwidth]{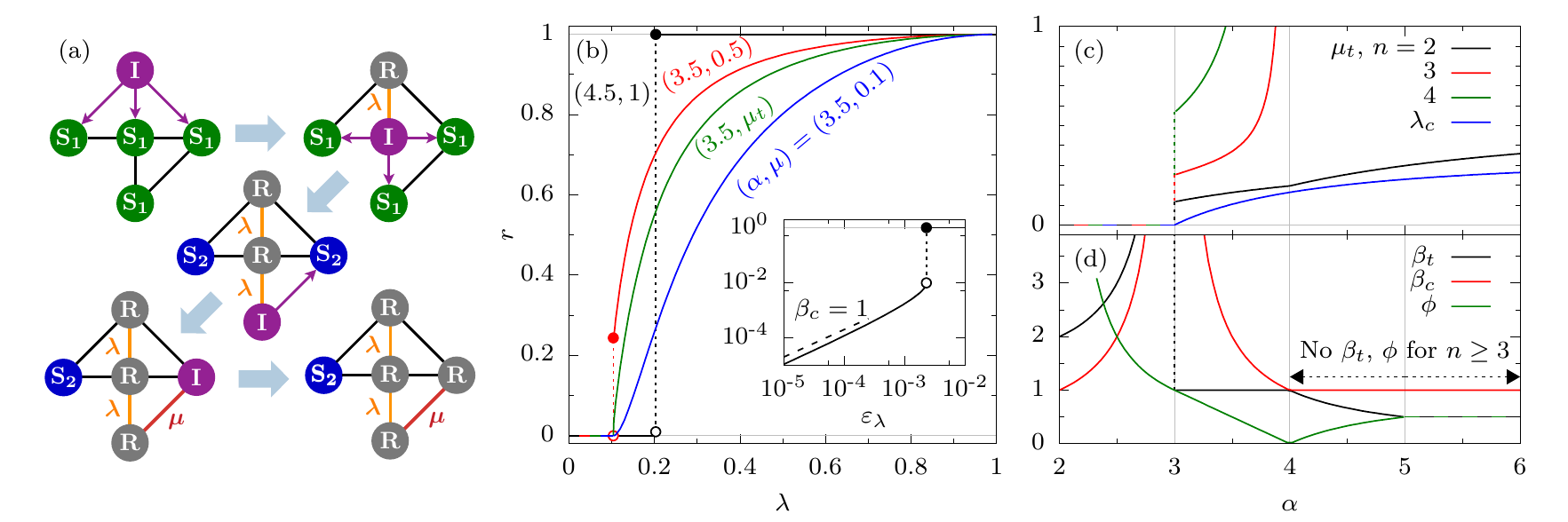}
\caption{\label{fig:fig1} (a) The GEP with $n = 3$ on a five-node network. Each thick arrow represents a time step. (b) Examples of the transitions of $r$ in the GEP with $n = 3$ on the SFNs. Inset: a magnified view of the double phase transition for $(\alpha,\mu) = (4.5,1)$. (c) The $\alpha$ dependence of the TCP $(\lambda_c,\mu_t)$ and (d) the scaling exponents in Table~\ref{tab}.
The SFNs in (b)--(d) have $k_m = 4$.}
\end{figure*}

{\em Dynamics.} In the GEP, a node can be susceptible ($\mathbf{S_1}$), weakened ($\mathbf{S_2}$), infected ($\mathbf{I}$), or removed ($\mathbf{R}$). All nodes are initially $\mathbf{S_1}$, except for one randomly chosen $\mathbf{I}$ node (the ``seed'') starting the spread. At each time step, a random $\mathbf{I}$ node attempts to spread the behavior to all of its $\mathbf{S_1}$- or $\mathbf{S_2}$-neighbors, each of the former (latter) with probability $\lambda$ ($\mu$). Upon success, the target becomes $\mathbf{I}$. A failed attempt does not affect the target unless it is the $(n-1)$-th attempt on the same $\mathbf{S_1}$ node, in which case the node becomes $\mathbf{S_2}$. After then, the chosen $\mathbf{I}$ node immediately deactivates and becomes $\mathbf{R}$, permanently removing itself from the dynamics. The process goes on until the network runs out of $\mathbf{I}$ nodes. The GEP with $n = 3$ on a five-node network is illustrated in Fig.~\hyperref[fig:fig1]{1(a)}.

{\em Substrate.} The GEP spreads on an ensemble of infinitely large random SFNs constrained by two conditions. First, the degree distribution obeys a power law $p_k = k^{-\alpha}/\zeta_{\alpha,k_m}$ for $k \ge k_m$ and $\alpha > 2$, where the generalized zeta function $\zeta_{s,v}$, defined as the analytic continuation of $\sum_{i=v}^\infty k^{-s}$ for $s \neq 1$, normalizes the distribution. The assumed range of $\alpha$ ensures that the mean degree $\langle k \rangle = \zeta_{\alpha-1,k_m}/\zeta_{\alpha,k_m}$ is finite. Second, there is no correlation between the degrees of adjacent nodes. Given these two conditions, one may assume that a node and each of its neighbors have mutually independent statistics, which makes the problem analytically tractable.

{\em Notations.} The final fraction of $\mathbf{R}$ nodes, denoted by $r$, quantifies the cascade size. The probability of a global cascade with $r > 0$ is denoted by $P_\infty$. The percolation transition from the phase with zero $r$ and $P_\infty$ to the phase with positive $r$ and $P_\infty$ occurs at $\lambda = \lambda_c$, and $r$ exhibits a continuous (discontinuous) transition at the point if $\mu \le \mu_t$ ($\mu > \mu_t$). The scaling behaviors near the TCP $(\lambda,\mu) = (\lambda_c,\mu_t)$ are characterized by three exponents $\beta_c$, $\beta_t$, and $\phi$, so that $P_\infty \sim \epsilon_\lambda^{\beta_c}$, $r \sim \epsilon_\lambda^{\beta_t}$, and $r \sim \epsilon_\mu^{\beta_t/\phi}$ with $\epsilon_\lambda \equiv (\lambda-\lambda_c)/\lambda_c$ and $\epsilon_\mu \equiv (\mu-\mu_t)/\mu_t$.

{\em Transition of $P_\infty$.} For the SFNs defined above, multiple spreading pathways rarely cross at the same node unless the cascade has already reached a finite fraction of the network. For this reason, $\mu$ is completely irrelevant to the transition from $P_\infty = 0$ to $P_\infty > 0$: only $\lambda$ controls the transition by a bond-percolation mechanism. Thus one can simply apply the theory of bond percolation on the random SFNs~\cite{CohenPRE2002} to obtain the transition point
\begin{align} \label{eq:lc}
\lambda_c = \begin{cases}
 \frac{\langle k \rangle}{\langle k(k-1) \rangle} = \frac{\zeta_{\alpha-1,k_m}}{\zeta_{\alpha-2,k_m}-\zeta_{\alpha-1,k_m}} &\text{for $\alpha > 3$,}\\
 0 &\text{for $2 < \alpha < 3$,}
 \end{cases}
\end{align}
which lies between $0$ and $1$ for sufficiently large $k_m$. The percolation theory~\cite{CohenPRE2002} also shows that the transition can only be continuous with the universal scaling behavior $P_\infty \sim \epsilon_\lambda^{\beta_c}$ for small positive $\epsilon_\lambda$, where the $\alpha$-dependent values of the critical exponent $\beta_c$ are listed in Table~\ref{tab}. Such equivalence has also been noted for the GEP~\cite{BizhaniPRE2012,KChungPRE2016} and cooperative contagions~\cite{WCaiNP2015,GrassbergerPRE2016,DLeeSciRep2017,WChoiPRE2017a,*WChoiPRE2017b,*WChoiPRE2018} on homogeneous networks.

\begin{table}[b]
\caption{\label{tab} Scaling exponents describing $P_\infty \sim \epsilon_\lambda^{\beta_c}$, $r \sim \epsilon_\lambda^{\beta_t}$, and $r \sim \epsilon_\mu^{\beta_t/\phi}$ of the GEP on the random SFNs near a TCP.}
\begin{tabularx}{\columnwidth}{*{4}{C}}
\hline\hline
& $\beta_c$ & $\beta_t$ & $\phi$ \\
\hline
$\alpha > 5$ & $1$ & $\frac{1}{2}$ & $\frac{1}{2}$ \\
$4 < \alpha < 5$ & $1$ & $\frac{1}{\alpha-3}$ & $\frac{\alpha-4}{\alpha-3}$ \\
$3 < \alpha < 4$ & $\frac{1}{\alpha-3}$ & $1$ & $4-\alpha$ \\
$2 < \alpha < 3$ & $\frac{1}{3-\alpha}$ & $\frac{4-\alpha}{3-\alpha}$ & $\frac{1}{\alpha-2}$ \\
\hline\hline
\end{tabularx}
\end{table}

{\em Analytic calculation of $r$.} In contrast to $P_\infty$, $r$ depends on $\mu$ as the crossing of spreading pathways is nonnegligible whenever $r > 0$. Here we present a calculation of the dependence based on a standard tree ansatz for random SFNs~\cite{CohenPRE2002}. For this aim, we consider the probability $q$ that a node at an end of a randomly chosen link is $\mathbf{R}$ after the spread has stopped. For simplicity, we assume $k_m \ge n-2$, which does not affect the main results. Then $q$ satisfies a self-consistency equation $q = f(q)$, where
\begin{align} \label{eq:q_map}
f(q) &\equiv 1 - \sum_{k=k_m}^{\infty} p'_k \Bigg[\sum_{m=0}^{k-1} \binom{k-1}{m}(1-\lambda)^{\min[m,n-1]}\nonumber\\
&\quad\times(1-\mu)^{\max[0,m-n+1]} q^m(1-q)^{k-1-m}\Bigg].
\end{align}
Each summand indexed by $m$ on the rhs accounts for the probability that the node has $m$ nodes among $k-1$ neighbors (excluding the neighbor at the other end of the randomly chosen link) trying to spread the behavior to it, all of which fail to do so. Note that $p'_k \equiv k p_k/\langle k \rangle$ is the degree distribution of a node at the end of a path, weighted by $k$ because higher-degree nodes are more likely to be connected. Once $q$ is known, we can calculate $r$ by
\begin{align} \label{eq:q_to_r}
r &= 1 - \sum_{k=k_m}^{\infty} p_k \Bigg[\sum_{m=0}^{k} \binom{k}{m}(1-\lambda)^{\min[m,n-1]}\nonumber\\
&\quad\times(1-\mu)^{\max[0,m-n+1]} q^m(1-q)^{k-m}\Bigg],
\end{align}
where $p_k$ appears instead of $p'_k$ because all nodes have equal weights regardless of $k$ in the definition of $r$. For any parameters, Eqs.~\eqref{eq:q_map} and \eqref{eq:q_to_r} provide an exact, albeit implicit, solution for $r$. Examples are shown in Fig.~\hyperref[fig:fig1]{1(b)} for the GEP with $n = 3$ on the SFNs with $k_m = 4$.

{\em Conditions for MOTs.} A MOT occurs at $\lambda = \lambda_c$ when it coincides with a discontinuous jump of $r$. Since Eq.~\eqref{eq:q_to_r} implies $r \simeq \langle k \rangle \lambda q$, the transitions of $r$ and $q$ should be of the same type. The latter are encoded in the small-$q$ expansion of Eq.~\eqref{eq:q_map}, which for noninteger $\alpha$ is given by (see Appendix~\ref{app:q_iter_exp} for the detailed derivation)
\begin{align} \label{eq:q_iter_exp}
&f(q) = \frac{\zeta_{\alpha-2,k_m}-\zeta_{\alpha-1,k_m}}{\zeta_{\alpha-1,k_m}}\lambda q\nonumber\\
& +\left(\frac{\zeta_{\alpha-3,k_m}-3\zeta_{\alpha-2,k_m}}{2\zeta_{\alpha-1,k_m}}+1\right)g_{2,n}(\lambda,\mu)\,q^2 \nonumber\\
& + \frac{\Gamma(2-\alpha)}{\zeta_{\alpha-1,k_m}}\,g_{\alpha-2,n}(\lambda,\mu)\,q^{\alpha-2} + O\!\left(q^{\min[3,\alpha-1]}\right),
\end{align}
where $\Gamma$ is the gamma function, and $g_{s,n}$ is defined as
\begin{align} \label{eq:g_def}
g_{s,n}(\lambda,\mu) \equiv &\left(\frac{1-\lambda}{1-\mu}\right)^{n-1}\Bigg\{-\mu^s + \sum_{m=0}^{n-2}\binom{m-1-s}{m} \nonumber\\
&\times(1-\mu)^m\left[1-\left(\frac{1-\mu}{1-\lambda}\right)^{n-1-m}\right]\Bigg\}.
\end{align}
Here $q^j$ with an integer $j$ corresponds to the contribution from $j$ neighbors, while $q^{\alpha-2}$ stems from the hubs. We note that the latter gets an extra factor of $\ln q$ for the special cases where $\alpha$ is an integer, which leads to some complications (see Appendix~\ref{app:q_iter_exp_int} for more details). The transition type is determined by whether $q = f(q)$ has a positive root at $\lambda = \lambda_c$, which in turn depends on the sign of $\bar f'' \equiv \lim_{q\downarrow 0} f''(q)$. If $\bar f'' > 0$ ($\bar f'' < 0$), a positive root exists (cannot exist), and the transition of $r$ is discontinuous (continuous). Applying this criterion to Eq.~\eqref{eq:q_iter_exp}, we find that the transition of $r$ is discontinuous (continuous) if $\mu > \mu_t$ ($\mu < \mu_t$), where $\mu_t \in [0,1]$ is a solution of
\begin{align} \label{eq:mt}
g_{\min[2,\alpha-2],n}(\lambda_c,\mu_t) = 0
\end{align}
for any noninteger $\alpha > 2$. In Fig.~\hyperref[fig:fig1]{1(c)}, we show examples of $\lambda_c$ and $\mu_t$ on the SFNs with $k_m = 4$ satisfying this equation. The solvability of Eq.~\eqref{eq:mt} has the following implications:

\noindent (i) If $\alpha > 4$, for $n = 2$ the solution is $\mu_t = \frac{\lambda_c}{1-\lambda_c}$, which depends on $\alpha$ only through $\lambda_c$. This is because the transition type is determined by the sign of $q^2$ in Eq.~\eqref{eq:q_iter_exp}, which is a two-neighbor effect. On the other hand, for $n \ge 3$ there is no solution because $g_{2,n}(\lambda_c,\mu_t) = -\lambda_c^2 < 0$; in other words, $\bar f'' < 0$ always holds, so the transition of $r$ is always continuous. Here $\mu$ comes into play only for three-or-more neighboring spreaders, so it cannot affect the sign of $q^2$.

\noindent (ii) If $3 < \alpha < 4$, Eq.~\eqref{eq:mt} is explicitly dependent on $\alpha$, reflecting the dominance of the hub-induced $q^{\alpha-2}$ term. Here the solution exists for any $n \ge 2$, because the convergence of many spreading pathways at the hubs facilitates a MOT even if $n$ is arbitrarily large. We note that $\mu_t$ obtained from Eq.~\eqref{eq:mt}, depending on $k_m$, can still be larger than $1$ and thus impossible to achieve, as shown for $k_m = 4$ in Fig.~\hyperref[fig:fig1]{1(c)}.

\noindent (iii) If $2 < \alpha < 3$, for any $n \ge 2$, $\mu_t = 0$ is the only solution. This captures $\lim_{\lambda\downarrow 0} r$ being positive (zero) for $\mu > 0$ ($\mu = 0$); in other words, there are so many spreading pathways crossing at the hubs that, regardless of $n$, synergistic spreading events by $\mu$ unaided by $\lambda$ can induce a global cascade. This regime is where the cascades of the GEP differ most significantly from those of cooperative contagions~\cite{CuiPRE2017}. In the latter, a node should first be infected by one contagion with probability $\lambda$ to experience a secondary infection with probability $\mu$, so $r = 0$ whenever $\lambda = 0$. In the GEP, even if $\lambda = 0$, a spreading event by $\mu$ can still occur because it only requires sufficiently many exposures to neighboring spreaders. This parallels the robust existence of a giant heterogeneous $k$-core with $k \ge 3$ on the same SFNs even in the limit where the fraction of removed nodes approaches unity~\cite{BaxterPRE2011}.

Based on these results, one can interpret the transition behaviors of the GEP with $n = 3$ on the SFNs with $k_m = 4$ illustrated in Fig.~\hyperref[fig:fig1]{1(b)}. For $\alpha = 3.5$, both continuous and discontinuous transitions of $r$ are possible at $\lambda_c \approx 0.104$ with the boundary at $\mu_t \approx 0.371$, whereas for $\alpha = 4.5$ (see the inset for a magnified view) $r$ undergoes a continuous transition belonging to the bond percolation universality class ($\beta_c = 1$) at $\lambda_c \approx 0.203$ even in the extreme case $\mu = 1$. Notably, there is a secondary discontinuous transition (marked by dotted vertical lines) at $\lambda > \lambda_c$, whose possibility is not excluded by our argument. This phenomenon seems to be related to the double phase transitions reported in \cite{BMinSciRep2018} and will be discussed in detail elsewhere~\cite{YBaekPrep}.

\begin{figure}
\includegraphics[width=\columnwidth]{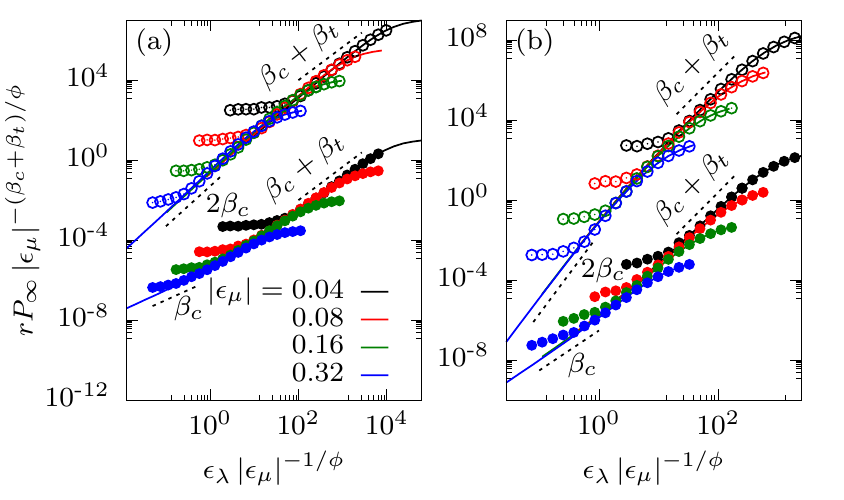}
\caption{\label{fig:fig2} The near-TCP crossover behaviors for $n = 2$ described by Eq.~\eqref{eq:rP_crossover}. The lines are obtained from the roots of Eq.~\eqref{eq:q_iter_exp}, and the symbols are simulation results obtained using $10^5$ SFNs with $N = 10^7$ and $k_m = 4$. The upper (lower) data correspond to the $\epsilon_\mu < 0$ ($\epsilon_\mu > 0$) regime with (a) $\alpha = 4.5$ and (b) $\alpha = 3.5$. See Fig.~\ref{fig:figS2} for the case $\alpha = 5.5$. To remove overlaps, all data for $\epsilon_\mu < 0$ have been divided by $10^6$. All plots use the same values of $|\epsilon_\mu|$.}
\vspace{-0.5cm}
\end{figure}

{\em Tricritical behaviors for $\alpha > 3$.} For small and positive $\epsilon_\lambda$, a Taylor expansion of Eq.~\eqref{eq:q_iter_exp} about $(\lambda,\mu)=(\lambda_c,\mu_t)$ yields
\begin{align} \label{eq:r_scaling}
	r \sim q \sim \begin{cases}
	(\epsilon_\lambda/|\epsilon_\mu|)^{\beta_c} &\text{if $|\epsilon_\mu| \gg \epsilon_\lambda^\phi$, $\epsilon_\mu < 0$,}\\
	\epsilon_\lambda^{\beta_t} &\text{if $|\epsilon_\mu| \ll \epsilon_\lambda^\phi$,}\\
	\epsilon_\mu^{\beta_t/\phi}	&\text{if $|\epsilon_\mu| \gg \epsilon_\lambda^\phi$, $\epsilon_\mu > 0$,}\\
	\end{cases}
\end{align}
where $\epsilon_\mu \equiv (\mu-\mu_t)/\mu_t$, the exponents $\beta_c$ and $\beta_t$ are shown in Table~\ref{tab} as well as Fig.~\hyperref[fig:fig1]{1(d)}, and $\phi = 1 - \beta_t/\beta_c$. The values of $\beta_t$ in this regime are in exact agreement with those reported in \cite{CuiPRE2017}. It is notable that the exponent $\phi$, which governs the crossover between different scaling regimes, exhibits nonmonotonic behaviors with the slope changing sign at $\alpha = 4$ [see Fig.~\hyperref[fig:fig1]{1(d)}]. This is yet another consequence of the fact that the hubs begin to drive the MOTs as $\alpha$ is decreased below $4$. 

To numerically verify the scaling exponents derived above, we present the scaling form for $r\,P_\infty$, which converges to the average fraction of $\mathbf{R}$ nodes, $\langle R \rangle/N$, readily obtained using random SFNs of $N$ nodes (see Appendix~\ref{app:net_gen} for more detail) in the $N\to\infty$ limit. The scaling form is given by
\begin{align} \label{eq:rP_crossover}
rP_\infty = \lim_{N \to \infty} \frac{\langle R \rangle}{N} =
|\epsilon_\mu|^{(\beta_t + \beta_c)/\phi} \, \mathcal{F}_\pm\!\left(\epsilon_\lambda\,|\epsilon_\mu|^{-1/\phi}\right),
\end{align}
where $\mathcal{F}_+$ ($\mathcal{F}_-$) is the scaling function for $\epsilon_\mu > 0$ ($\epsilon_\mu < 0$). As shown in Fig.~\ref{fig:fig2}, there is a good agreement between the theory and the numerics, despite deviations due to finite-size effects for small $|\epsilon_\lambda|$ and $|\epsilon_\mu|$ (see Fig.~\ref{fig:figS3} for a closer comparison between theory and numerics). 

\begin{figure}
\includegraphics[width=\columnwidth]{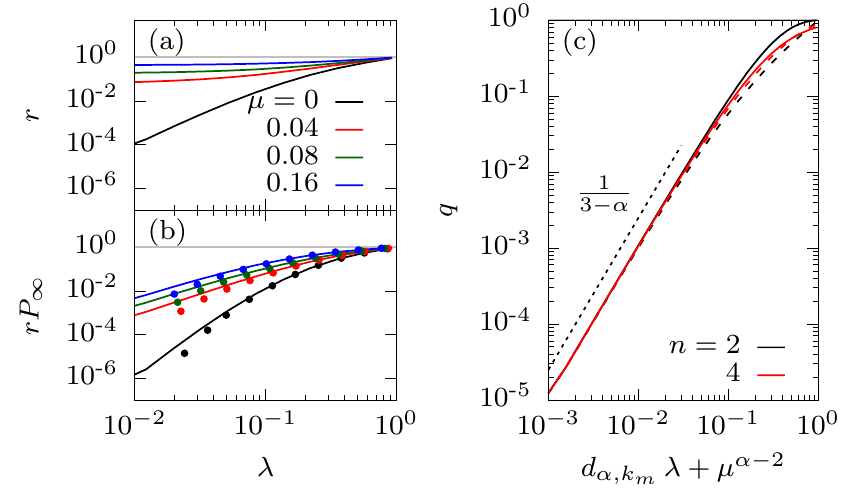}
\caption{\label{fig:fig3} (a) Scaling behaviors of the cascade size $r$ on the SFNs with $\alpha = 2.5$ and $k_m = 4$. (b) Comparison between the asymptotic values of $r P_\infty$ (solid lines) predicted by the roots of Eq.~\eqref{eq:q_iter_exp} and the corresponding finite-size observable $\langle R \rangle/N$ (symbols) numerically obtained from $10^5$ networks with $N = 10^6$. Both (a) and (b) use $n = 2$ and the same values of $\mu$. (c) Universal scaling form of $q$ with respect to $d_{\alpha,k_m}\lambda + \mu^{\alpha-2}$, as predicted by Eq.~\eqref{eq:a25_scaling}. The solid (dashed) lines correspond to $\lambda = 0$ ($\mu = 0$).}
\vspace{-0.5cm}
\end{figure}

{\em Scaling behaviors for $2 < \alpha < 3$.} As discussed above and illustrated in Figs.~\hyperref[fig:fig3]{3(a)} and \hyperref[fig:fig3]{3(b)} (the latter providing a numerical verification of the tree ansatz, whose rigorous justification remains an open mathematical problem due to a diverging number of short loops~\cite{BianconiJSM2005}), $\lambda_c = \mu_t = 0$ holds in this regime. Due to the absence of the phase of localized cascades, it would be misleading to call the point a TCP; however, one can still identify universal scaling behaviors and the crossover between them from the leading-order terms of Eq.~\eqref{eq:q_iter_exp}, identifying new scaling exponents previously unreported. We obtain
\begin{align} \label{eq:a25_scaling}
q \sim \left(d_{\alpha,k_m}\lambda + \mu^{\alpha-2}\right)^{1/(3-\alpha)}
\end{align}
with a coefficient $d_{\alpha,k_m} > 0$ determined by $\alpha$ and $k_m$, as illustrated in Fig.~\hyperref[fig:fig3]{3(c)}. For $\mu = 0$, the above equation and $r \sim \lambda q$ from Eq.~\eqref{eq:q_to_r} implies $r \sim \lambda^{\beta_t}$ with $\beta_t = \frac{4-\alpha}{3-\alpha}$. Moreover, since the positive limiting values of $q$ and $r$ as $\lambda$ decreases to zero become clear only for $\mu \gg \lambda^\frac{1}{\alpha-2}$, we can also write $\phi = \frac{1}{2-\alpha}$ to describe the crossover. The behaviors of $\beta_t$ and $\phi$ for $2 < \alpha < 3$ shown in Table~\ref{tab} and Fig.~\hyperref[fig:fig1]{1(c)} should be understood in this vein.

{\em Summary.} We examined the effects of the degree exponent $\alpha$ on the percolation transitions of the GEP on uncorrelated random SFNs. All analytical results, based on the tree ansatz~\eqref{eq:q_map}, are in good agreement with the numerics beyond the regime of strong finite-size effects. It is found that the hub-driven MOTs occur only for $\alpha < 4$. In particular, for $2 < \alpha < 3$, we identified new transition behaviors stemming from the convergence of loops at the hubs. These imply that the spread of behavior and cooperative contagions~\cite{CuiPRE2017} belong to different universality classes on typical social networks. Our results reveal fundamental principles underlying the formation of compact cultural subgroups fostered by the fat-tailed degree distribution of social networks. Interesting topics for future studies include the conditions for double phase transitions, the nature of finite-size effects, and connections to MOTs and TCPs reported in other percolation models~\cite{CellaiPRL2011,AraujoPRL2011}.

{\em Acknowledgments.} This research was supported by Basic Science Research Program through the National Research Foundation of Korea (NRF)~(KR) [NRF-2017R1D1A3A03000578 (M.H.) and NRF-2017R1A2B3006930 (H.J.)]. Y.B. is supported in part by the European Research Council under the Horizon 2020 Programme, ERC Grant Agreement No. 740269. We also thank Peter Grassberger for helpful comments on mixed-order transitions in DNA unzipping as well as the references.

\onecolumngrid
\appendix

\setcounter{figure}{0}
\setcounter{table}{0}
\makeatletter
\renewcommand{\thefigure}{S\arabic{figure}}
\renewcommand{\thetable}{S\arabic{table}}

\begin{spacing}{1.3}

\section{Generation of scale-free networks} \label{app:net_gen}

In our simulations of the generalized epidemic process (GEP), we randomly generated the scale-free networks (SFNs) according to the following three-step scheme.

{\bf Step 1.} Depending on the value of $\alpha$, fix the maximum degree as
\begin{align}
k_\text{max} = \begin{cases}
 	N - 1 &\text{if $\alpha \ge 3$,}\\
 	\lfloor\sqrt{N}\rfloor &\text{if $2 < \alpha < 3$.}
 \end{cases}
\end{align}
This ensures that the degrees of adjacent nodes are uncorrelated~\cite{CatanzaroPRE2005}.

{\bf Step 2.} Given the degree distribution
\begin{equation}
p_k = \frac{k^{-\alpha}}{\sum_{k' = k_m}^{k_\text{max}} k'^{-\alpha}}, \label{eq:deg_dist}
\end{equation}
generate a degree sequence {\em deterministically} so that the number of nodes with degree $k$, denoted by $N_k$, satisfies
\begin{equation}
\left \lfloor N \sum_{k'>k}p_{k'} \right \rfloor = \sum_{k'>k} N_{k'},
\end{equation}
for every integer $k \in [k_m,k_\text{max}]$. This method, used in \cite{NohPRE2009}, reduces the noise stemming from the sample-to-sample fluctuations of the degree sequence at finite $N$.

{\bf Step 3.} Randomly connect the nodes according to the given degree sequence, avoiding the creation of self-loops and multiple links between the same pair of nodes.

\section{Derivation of Eq.~\eqref{eq:q_iter_exp}}
\label{app:q_iter_exp}

\begin{table}
\caption{\label{tab:stirling} Unsigned Stirling numbers of the first kind $\stirling{j}{i}$ for small non-negative integers $j$ and $i$.}
\begin{tabularx}{\columnwidth}{C|*{8}{C}}
\hline\hline
\diagbox[width=\dimexpr\linewidth+2\tabcolsep]{$~~~~j$}{$i~~~~$} & $0$ & $1$ & $2$ & $3$ & $4$ & $5$ & $6$ & $7$ \\
\hline
$0$ & $1$ & -- & -- & -- & -- & -- & -- & --  \\
$1$ & $0$ & $1$ & -- & -- & -- & -- & -- & -- \\
$2$ & $0$ & $1$ & $1$ & -- & -- & -- & -- & -- \\
$3$ & $0$ & $2$ & $3$ & $1$ & -- & -- & -- & -- \\
$4$ & $0$ & $6$ & $11$ & $6$ & $1$ & -- & -- & -- \\
$5$ & $0$ & $24$ & $50$ & $35$ & $10$ & $1$ & -- & -- \\
$6$ & $0$ & $120$ & $274$ & $225$ & $85$ & $15$ & $1$ & -- \\
$7$ & $0$ & $720$ & $1764$ & $1624$ & $735$ & $175$ & $21$ & $1$ \\
\hline\hline
\end{tabularx}
\end{table}

We first rewrite Eq.~\eqref{eq:q_map} as
\begin{align} \label{eq:q_iter_binom}
f(q) &= 1-\sum_{k=k_m}^{\infty} p'_k \Bigg[\sum_{m=0}^{n-1}\binom{k-1}{m}(1-\lambda)^m q^m (1-q)^{k-1-m} + \sum_{m=n}^{k-1} \binom{k-1}{m}(1-\lambda)^{n-1}(1-\mu)^{m-n+1}q^m(1-q)^{k-1-m}\Bigg] \nonumber\\
&= 1-\sum_{k=k_m}^{\infty} p'_k \Bigg\{\left(\frac{1-\lambda}{1-\mu}\right)^{n-1}(1-\mu q)^{k-1} + \sum_{m=0}^{n-2}\binom{k-1}{m}(1-\lambda)^m\left[1-\left(\frac{1-\lambda}{1-\mu}\right)^{n-m-1}\right]q^m(1-q)^{k-1-m}\Bigg\},
\end{align}
whose validity can be easily shown by the binomial expansion of $(1-\mu q)^{k-1}$. Using a notation for the {\em Lerch transcendent}
\begin{align} \label{eq:lerch_def}
\Phi_{s,v}(z) \equiv \sum_{i=0}^\infty \frac{z^i}{(v+i)^s},
\end{align}
we can calculate the summations over $k$ in Eq.~\eqref{eq:q_iter_binom} to obtain
\begin{align} \label{eq:q_iter}
f(q) &= 1 - \frac{1}{\zeta_{\alpha-1,k_m}}\left(\frac{1-\lambda}{1-\mu}\right)^{n-1}(1-\mu q)^{k_m-1}\Phi_{\alpha-1,k_m}(1-\mu q)\nonumber\\
&\quad - \frac{1}{\zeta_{\alpha-1,k_m}}\sum_{m=0}^{n-2} \frac{(1-\lambda)^m}{m!}\left[1-\left(\frac{1-\lambda}{1-\mu}\right)^{n-m-1}\right](-q)^m\frac{d^m}{dq^m}\left[(1-q)^{k_m-1}\Phi_{\alpha-1,k_m}(1-q)\right].
\end{align}
In order to expand the rhs of Eq.~\eqref{eq:q_iter} with respect to $q$, we note that the Lerch transcendent has a series expansion~\cite{Bateman1953}
\begin{align} \label{eq:lerch_exp_log}
\Phi_{s,v}(z) = z^{-v}\sum_{i=0}^\infty \zeta_{s-i,v} \frac{(\ln z)^i}{i!} + z^{-v}\Gamma(1-s)(-\ln z)^{s-1}
\end{align}
for any complex $z$ with $|\ln z| < 2\pi$ and for real numbers $s$ and $v$ satisfying $s \neq 1,\,2,\,3,\dotsc$ and $v \neq 0,-1,-2,\dots$. Taking advantage of the generating function
\begin{align} \label{eq:stirling_gen}
[\ln(1-x)]^i = (-1)^i\cdot i!\cdot\sum_{j=i}^\infty\stirling{j}{i}\frac{x^j}{j!}
\end{align}
for the {\em unsigned Stirling numbers of the first kind} $\stirling{j}{i}$ (whose values for small $j$ and $i$ are listed in Table~\ref{tab:stirling}), we can derive a useful relation
\begin{align}
\frac{[\ln(1-x)]^i}{1-x} = -\frac{1}{i+1}\frac{d}{dx}[\ln(1-x)]^{i+1}
= (-1)^i\cdot i! \cdot \sum_{j=i+1}^\infty \stirling{j}{i+1}\frac{x^{j-1}}{(j-1)!}.
\end{align}
This in turn can be used to rewrite Eq.~\eqref{eq:lerch_exp_log} in a more convenient form
\begin{align}
&(1-x)^{v-1}\,\Phi_{s,v}(1-x) = \sum_{j=1}^\infty \left\{\sum_{i=0}^{j-1} (-1)^i\stirling{j}{i+1}\zeta_{s-i,v}\right\}\frac{x^{j-1}}{(j-1)!} + \frac{\Gamma(1-s)}{\zeta_{s,v}} x^{s-1}\left[1 + O\!\left(x\right)\right] \nonumber\\
&\quad = \sum_{j=0}^\infty \left\{\sum_{i=1}^{j+1} (-1)^{i+1}\stirling{j+1}{i}\zeta_{s-i+1,v}\right\}\frac{x^j}{j!} + \frac{\Gamma(1-s)}{\zeta_{s,v}} \left[x^{s-1} + O\!\left(x^s\right)\right],
\end{align}
where the second equality is obtained by the change of variables $j \to j+1$ and $i \to i-1$. Using the above expansion in Eq.~\eqref{eq:q_iter}, a tedious but straightforward calculation yields
\begin{align}
&f(q) = \sum_{j=1}^{\infty}\left\{\sum_{i=1}^{j+1} \frac{(-1)^{i+1}}{j!}\stirling{j+1}{i}\frac{\zeta_{\alpha-i,k_m}}{\zeta_{\alpha-1,k_m}}\right\}\left(\frac{1-\lambda}{1-\mu}\right)^{n-1}
\left\{\sum_{m=0}^{n-2}\binom{m-1-j}{m}(1-\mu)^m\left[1-\left(\frac{1-\mu}{1-\lambda}\right)^{n-1-m}\right]-\mu^j\right\}q^j\nonumber\\
&\quad + \frac{\Gamma(2-\alpha)}{\zeta_{\alpha-1,k_m}}\left(\frac{1-\lambda}{1-\mu}\right)^{n-1}
\left\{\sum_{m=0}^{n-2}\binom{m+1-\alpha}{m}(1-\mu)^m\left[1-\left(\frac{1-\mu}{1-\lambda}\right)^{n-1-m}\right]-\mu^{\alpha-2}\right\}\left[q^{\alpha-2}+O\!\left(q^{\alpha-1}\right)\right],
\end{align}
where $\binom{m'}{m}$ is a generalized Binomial coefficient defined as
\begin{align}
\binom{m'}{m} \equiv \frac{m'(m'-1)\cdots(m'-m+1)}{m!}
\end{align}
for any integer $m'$ and a non-negative integer $m$. The definition implies $\binom{m'}{m} = (-1)^m\binom{|m'|+m-1}{m}$ for any negative $m'$ and $\binom{m'}{m} = 0$ whenever $m > m' \ge 0$. Using these properties and Table~\ref{tab:stirling}, the order $q$ component of $f(q)$ is given by
\begin{align}
&\frac{\zeta_{\alpha-1,k_m}-\zeta_{\alpha-2,k_m}}{\zeta_{\alpha-1,k_m}}\left(\frac{1-\lambda}{1-\mu}\right)^{n-1}\left\{\sum_{m=0}^{n-2}\binom{m-2}{m}(1-\mu)^m\left[1-\left(\frac{1-\mu}{1-\lambda}\right)^{n-1-m}\right]-\mu\right\}q\nonumber\\
&\quad= \frac{\zeta_{\alpha-1,k_m}-\zeta_{\alpha-2,k_m}}{\zeta_{\alpha-1,k_m}}\left(\frac{1-\lambda}{1-\mu}\right)^{n-1}\left\{\left[1-\left(\frac{1-\mu}{1-\lambda}\right)^{n-1}\right]-(1-\mu)\left[1-\left(\frac{1-\mu}{1-\lambda}\right)^{n-2}\right]\theta_{n-3}-\mu\right\}q\nonumber\\
&\quad = \frac{\zeta_{\alpha-2,k_m}-\zeta_{\alpha-1,k_m}}{\zeta_{\alpha-1,k_m}}\lambda q,
\end{align}
where $\theta_m = 1$ ($\theta_m = 0$) for any integer $m \ge 0$ ($m < 0$). 
Then Eq.~\eqref{eq:q_iter_exp} is obtained by defining $g_{s,n}(\lambda,\mu)$ as in Eq.~\eqref{eq:g_def}.

\section{Phase transitions at integer degree exponents}
\label{app:q_iter_exp_int}

If the degree exponent $\alpha$ is an integer, the epidemic outbreaks and their associated critical phenomena are governed by the behavior of $\Phi_{s,v}(z)$ near $z = 1$ for a positive integer $s$. The relevant series expansion is given by~\cite{Bateman1953}
\begin{align} \label{eq:lerch_exp_nonint_s}
\Phi_{s,v}(z) \equiv z^{-v}\sum_{n=0}^\infty \tilde{\zeta}_{s-n,v} \frac{(\ln z)^n}{n!} + z^{-v}\left[\psi(s)-\psi(v)-\ln(-\ln z)\right]\frac{(\ln z)^{s-1}}{(s-1)!}
\end{align}
for $|\ln z| < 2\pi$ and $v \neq 0,-1,-2,\dots$, where we have introduced the notations
\begin{align}
\tilde{\zeta}_{s,v} \equiv \begin{cases}
	\zeta_{s,v} &\text{if $s \ge 2$,}\\
	0 &\text{if $s = 1$}
 \end{cases}
\end{align}
and $\psi(s) \equiv \Gamma'(s)/\Gamma(s)$ for the digamma function. Using Eq.~\eqref{eq:stirling_gen}, we can recast the above expansion into a more convenient form
\begin{align}
(1-x)^{v-1}\,\Phi_{s,v}(1-x) &= \sum_{j=0}^\infty \left\{\sum_{i=1}^{j+1} (-1)^{i+1}\stirling{j+1}{i}\tilde{\zeta}_{s-i+1,v}\right\}\frac{x^j}{j!} -\frac{(-1)^{s-1}}{(s-1)!}\left\{x^{s-1}\ln x + \left[\psi(v) - \psi(s)\right]x^{s-1}\right\} + O\!\left(x^s\right).
\end{align}
Based on this formula, we can expand the rhs of Eq.~\eqref{eq:q_iter} as
\begin{align} \label{eq:q_iter_exp_int}
f(q) &= \sum_{j=1}^{\infty}\left\{\sum_{i=1}^{j+1} \frac{(-1)^{i+1}}{j!}\stirling{j+1}{i}\frac{\tilde{\zeta}_{\alpha-i,k_m}}{\zeta_{\alpha-1,k_m}}\right\} g_{j,n}(\lambda,\mu)\, q^j - \frac{(-1)^{\alpha-2}}{\zeta_{\alpha-1,k_m}(\alpha-2)!} g_{\alpha-2,n}(\lambda,\mu)\,q^{\alpha-2}\ln q\nonumber\\
&\quad - \frac{(-1)^{\alpha-2}}{\zeta_{\alpha-1,k_m}(\alpha-2)!}\Bigg[ \left[\psi(k_m)-\psi(\alpha-1)\right]g_{\alpha-2,n}(\lambda,\mu)-\left(\frac{1-\lambda}{1-\mu}\right)^{n-1}\Bigg\{\mu^{\alpha-2}\ln \mu \nonumber\\
&\quad - \sum_{m=0}^{\min[\alpha,n]-2}(1-\mu)^m\left[1-\left(\frac{1-\mu}{1-\lambda}\right)^{n-1-m}\right]\binom{m+1-\alpha}{m}\left[\psi(\alpha-1)-\psi(\alpha-1-m)\right]\Bigg\}\Bigg]q^{\alpha-2} + O\!\left(q^{\alpha-1}\ln q\right),
\end{align}
where we have used $g_{s,n}(\lambda,\mu)$ defined in Eq.~\eqref{eq:g_def}. The main difference between Eq.~\eqref{eq:q_iter_exp} and Eq.~\eqref{eq:q_iter_exp_int} lies in the presence of $q^{\alpha-2}\ln q$ in the latter, which is always lower-order than $q^{\alpha-2}$. If $\alpha > 5$, the term is simply irrelevant to epidemic outbreaks. If $\alpha \in \{3,4,5\}$, the logarithmic correction has nontrivial effects on the transition behaviors, as discussed case by case below (see Table~\ref{tab:integ} for a summary).

\noindent {\bf Case of $\bm{\alpha = 5}$:} the lowest-order terms of Eq.~\eqref{eq:q_iter_exp_int} are given by
\begin{align} \label{eq:q_iter_exp_a5}
f(q) &= \frac{\zeta_{3,k_m}-\zeta_{4,k_m}}{\zeta_{4,k_m}}\lambda q + \frac{\zeta_{2,k_m}-3\zeta_{3,k_m}+2\zeta_{4,k_m}}{2\zeta_{4,k_m}} g_{2,n}(\lambda,\mu)\, q^2 + \frac{g_{3,n}(\lambda,\mu)}{6\zeta_{4,k_m}}\,q^3\ln q + O\!\left(q^3\right),
\end{align}
whose form is similar to the corresponding recursive relation for a non-integer $\alpha > 4$. Based on the same arguments described in the main text, the epidemic threshold is obtained as $\lambda_c = \zeta_{4,k_m}/(\zeta_{3,k_m}-\zeta_{4,k_m})$, and the tricritical point (TCP) satisfies $g_{2,n}(\lambda_c,\mu_t) = 0$, which has a physical solution $\mu_t = \lambda_c/(1-\lambda_c) \in (0,1)$ for $n = 2$ and sufficiently large $k_m$. Near the TCP, we can approximate the above equation as
\begin{align}
0 \simeq \epsilon_\lambda q + c_{\alpha,k_m} \epsilon_\mu q^2 - c'_{\alpha,k_m} q^3 |\ln q|,
\end{align}
where $c_{\alpha,k_m}$ and $c'_{\alpha,k_m}$ are positive coefficients. Thus the behavior of the outbreak size in this regime satisfies
\begin{align}
r \sim q \sim \begin{cases}
\epsilon_\lambda/|\epsilon_\mu| &\text{if $\epsilon_\mu < 0$, $|\epsilon_\mu| \gg |\epsilon_\lambda \ln \epsilon_\lambda|^{1/2}$,}\\
|\epsilon_\lambda/\ln\epsilon_\lambda|^{1/2} &\text{if $|\epsilon_\mu| \ll |\epsilon_\lambda \ln \epsilon_\lambda|^{1/2}$,}\\
\epsilon_\mu/|\ln \epsilon_\mu| &\text{if $\epsilon_\mu > 0$, $|\epsilon_\mu| \gg |\epsilon_\lambda \ln \epsilon_\lambda|^{1/2}$.}
\end{cases}
\end{align}

\noindent {\bf Case of $\bm{\alpha = 4}$:} the lowest-order terms of Eq.~\eqref{eq:q_iter_exp_int} are obtained as
\begin{align} \label{eq:q_iter_exp_a4}
f(q) &= \frac{\zeta_{2,k_m}-\zeta_{3,k_m}}{\zeta_{3,k_m}}\lambda q - \frac{g_{2,n}(\lambda,\mu)}{2\zeta_{3,k_m}}\,q^2\ln q \nonumber\\
&\quad - \frac{1}{2\zeta_{3,k_m}}\Bigg[ \left[3\zeta_{2,k_m}-2\zeta_{3,k_m} + \psi(k_m)-\psi(3)\right]g_{2,n}(\lambda,\mu)-\left(\frac{1-\lambda}{1-\mu}\right)^{n-1}\Bigg\{\mu^{2}\ln \mu \nonumber\\
&\quad - \sum_{m=0}^{\min[4,n]-2}(1-\mu)^m\left[1-\left(\frac{1-\mu}{1-\lambda}\right)^{n-1-m}\right]\binom{m-3}{m}\left[\psi(3)-\psi(3-m)\right]\Bigg\}\Bigg]q^2 + O\!\left(q^3\ln q\right),
\end{align}
which implies that the epidemic threshold is at $\lambda_c = \zeta_{3,k_m}/(\zeta_{2,k_m}-\zeta_{3,k_m})$ and that the TCP satisfies $g_{2,n}(\lambda_c,\mu_t) = 0$. As was the case for $\alpha > 4$, the TCP exists only for $n = 2$ and sufficiently large $k_m$. The near-TCP properties are described by
\begin{align}
0 \simeq \epsilon_\lambda q + c_{\alpha,k_m} \epsilon_\mu q^2 |\ln q| - c'_{\alpha,k_m} q^2,
\end{align}
for positive coefficients $c_{\alpha,k_m}$ and $c'_{\alpha,k_m}$. Thus the outbreak size in this regime obeys
\begin{align}
r \sim q \sim \begin{cases}
\epsilon_\lambda/|\epsilon_\mu \ln (\epsilon_\lambda/|\epsilon_\mu|)| &\text{if $\epsilon_\mu < 0$, $|\epsilon_\mu| \gg |\ln \epsilon_\lambda|^{-1}$,}\\
\epsilon_\lambda &\text{if $|\epsilon_\mu| \ll |\ln \epsilon_\lambda|^{-1}$,}\\
\mathrm{e}^{-c'_{\alpha,k_m}/\left(c_{\alpha,k_m}\epsilon_\mu\right)} &\text{if $\epsilon_\mu > 0$, $|\epsilon_\mu| \gg |\ln \epsilon_\lambda|^{-1}$.}
\end{cases}
\end{align}

\noindent {\bf Case of $\bm{\alpha = 3}$:} the lowest-order terms of Eq.~\eqref{eq:q_iter_exp_int} are given by
\begin{align} \label{eq:q_iter_exp_a3}
f(q) &= -\frac{1}{\zeta_{2,k_m}}\lambda q \ln q - \frac{1}{\zeta_{2,k_m}}\Bigg[ \left[\zeta_{2,k_m}+\psi(k_m)-\psi(2)\right]\lambda+\left(\frac{1-\lambda}{1-\mu}\right)^{n-1}\Bigg\{\mu\ln \mu \nonumber\\
&\quad - \sum_{m=0}^{\min[3,n]-2}(1-\mu)^m\left[1-\left(\frac{1-\mu}{1-\lambda}\right)^{n-1-m}\right]\binom{m-2}{m}\left[\psi(2)-\psi(2-m)\right]\Bigg\}\Bigg]q + O\!\left(q^{2}\right).
\end{align}
At the vanishing epidemic threshold ($\lambda_c = 0$), $q = f(q)$ has (cannot have) a positive root if the sign of the $q$ term on the rhs is positive (negative). Thus $\mu_t$ is given by
\begin{align} \label{eq:mu_t_a3}
\mu_t\ln \mu_t = \sum_{m=0}^{\min[3,n]-2}(1-\mu_t)^m\left[1-\left(1-\mu_t\right)^{n-1-m}\right]\binom{m-2}{m}\left[\psi(2)-\psi(2-m)\right].
\end{align}
We note that $\mu_t$ obtained from the above equation is in general not equal to $\lim_{\alpha\downarrow 3}\mu_t$ obtained from Eq.~\eqref{eq:mt}. If $\mu < \mu_t$, the transition behaviors are described by the approximate formula
\begin{align}
0 \simeq c_{\alpha,k_m} \lambda q |\ln q| + \left(c'_{\alpha,k_m}\epsilon_\mu-c''_{\alpha,k_m}\lambda\right) q,
\end{align}
where $c_{\alpha,k_m}$, $c'_{\alpha,k_m}$, and $c''_{\alpha,k_m}$ are positive coefficients. In this case, the outbreak size satisfies
\begin{align}
r \sim \lambda q \sim \lambda\,\mathrm{e}^{\left(c'_{\alpha,k_m}\epsilon_\mu-c''_{\alpha,k_m}\lambda\right)/\left(c_{\alpha,k_m}\lambda\right)}.
\end{align}
As $\epsilon_\mu$ approaches zero so that $|\epsilon_\mu| \ll \lambda$ (which can be represented as $\phi = 1$), $r$ abruptly becomes nonzero for an arbitrary positive value of $\lambda$. In contrast to the other cases, here $r$ can be already nonzero at $\lambda = \lambda_c$ and $\mu = \mu_t$ in a manner analogous to a discontinuous transition.

\begin{table}
\caption{\label{tab:integ} Scaling exponents describing tricritical properties of the GEP (if TCPs exist) on random SFNs for integer degree exponents $\alpha$.}
\begin{tabularx}{\columnwidth}{*{4}{C}}
\hline\hline
& $P_\infty \sim \epsilon_\lambda^{\beta_c}$ & $r \sim \epsilon_\lambda^{\beta_t}$ & $\epsilon_\mu \sim \epsilon_\lambda^\phi$ \\
\hline
$\alpha = 5$ & $\epsilon_\lambda$ & $|\epsilon_\lambda/\ln \epsilon_\lambda|^{1/2}$ & $|\epsilon_\lambda \ln \epsilon_\lambda |^{1/2}$ \\
$\alpha = 4$ & $|\epsilon_\lambda/\ln\epsilon_\lambda|$ & $\epsilon_\lambda$ & $|\ln \epsilon_\lambda|^{-1}$ \\
$\alpha = 3$ & $\lambda e^{-c/\lambda}$ & $\lambda^0$ & $\lambda$ \\
\hline\hline
\end{tabularx}
\end{table}

\section{Illustrations of actual outbreaks}

\begin{figure}
\includegraphics[width=\columnwidth]{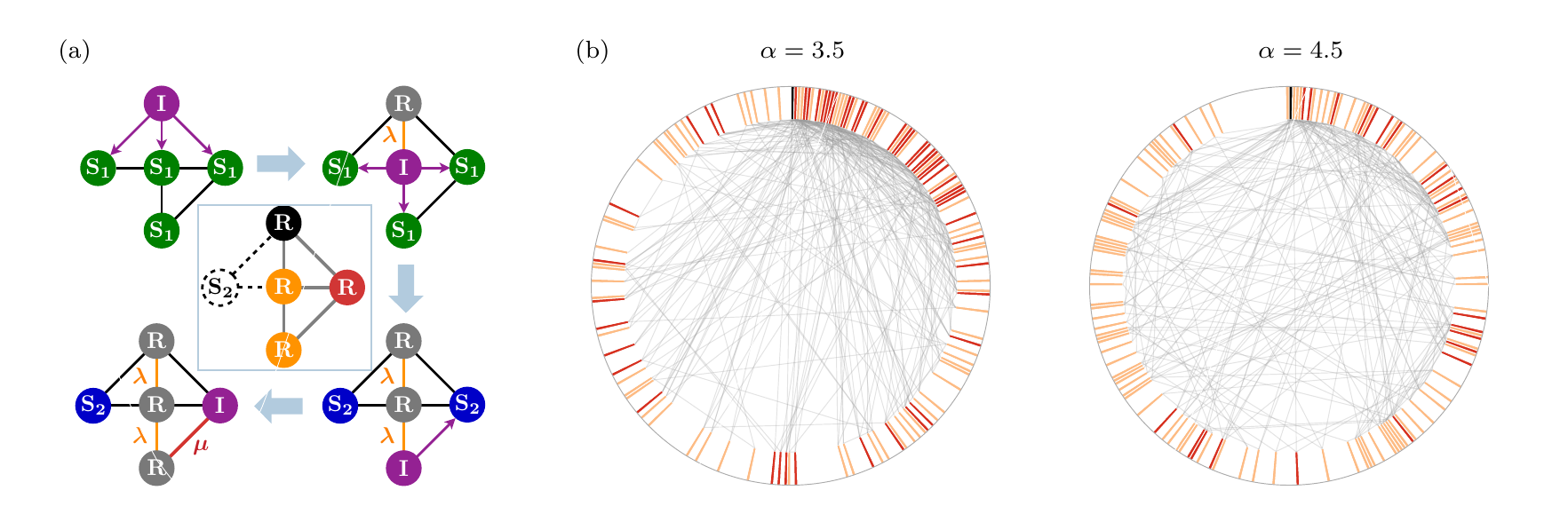}
\caption{\label{fig:figS1} Examples of the GEP with $n = 3$. (a) Entire dynamics on a five-node network. Each thick arrow represents a time step. Central box: in the final state, the seed is colored black, the nodes infected with probability $\lambda$ ($\mu$) are colored orange (red), and only the links connecting the infected nodes are shown. (b) Examples of the final state of the GEP on the SFNs with $k_m = 4$ at $\lambda = \lambda_c$, and $\mu = 0.5$. The rods (both colored and white) on the boundary correspond to the nodes, aligned clockwise in the order of decreasing degree. Only the infected nodes and their mutual links are shown according to the color scheme shown in (a). Here the seed is located at the node of the highest degree (the black rod).}
\end{figure}

The importance of hubs in the MOTs for $3 < \alpha < 4$ is more directly illustrated in Fig.~\ref{fig:figS1}. Using the color scheme described in Fig.~\hyperref[fig:figS1]{S1(a)}, each circular diagram of Fig.~\hyperref[fig:figS1]{S1(b)} shows the final state of the GEP with $n = 3$ at $\lambda = \lambda_c$ and $\mu = 0.5$ on the random SFNs with $N = 360$ nodes and $k_m = 4$. More specifically, each rod on the periphery corresponds to a node, aligned clockwise in the order of decreasing degree (nodes of equal degree are randomly ordered). The seed node (chosen to be the node of the highest degree) is black, the nodes infected in the $\mathbf{S_1}$-state are orange, and those infected in the $\mathbf{S_2}$-state are red. The uninfected nodes are left as vacancies. The links are drawn with grey lines only if they connect two infected neighbors. By comparing these two examples of epidemic outbreaks at $\alpha = 3.5$ and $4.5$, it is clear that the $\mathbf{S_2} \to \mathbf{I}$ infections (red nodes) are especially frequent among the high-degree nodes in the case of $\alpha = 3.5$. This reflects the dominant role played by the hubs in the system-wide avalanche for $3 < \alpha < 4$ (note that $\mu = 0.5 > \mu_t \approx 0.371$ in this case). In contrast, for $\alpha = 4.5$, the high cooperation threshold $n = 3$ and the dominance of two-neighbor effects reduce the significance of cooperative infections among the hubs at the transition, which is bound to be purely continuous. Consequently, the nodes infected by the cooperative mechanism are more evenly distributed among different degrees in the latter case.

\section{Near-TCP crossover for $\alpha = 5.5$}

\begin{figure}[h]
\includegraphics[width=0.5\columnwidth]{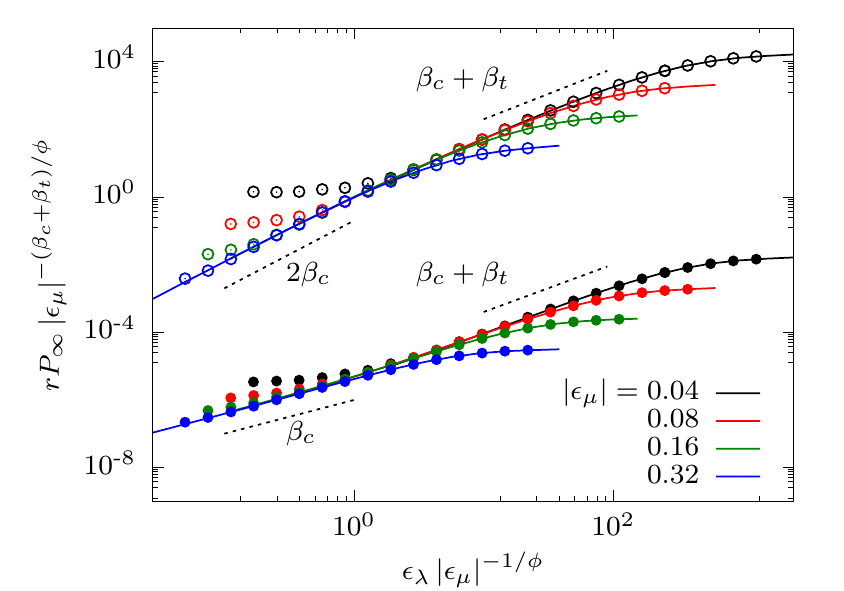}
\caption{\label{fig:figS2} The near-TCP crossover behaviors for $\alpha = 5.5$ and $n = 2$ described by Eq.~\eqref{eq:rP_crossover}. The lines are obtained from the roots of Eq.~\eqref{eq:q_iter_exp}, and the symbols are simulation results obtained using $10^5$ SFNs with $N = 10^7$ and $k_m = 4$. The upper (lower) data correspond to the $\epsilon_\mu < 0$ ($\epsilon_\mu > 0$) regime. To remove overlaps, all data for $\epsilon_\mu < 0$ have been divided by $10^6$.}
\end{figure}

In Fig.~\ref{fig:figS2}, we show the near-TCP crossover behaviors for the GEP with $n = 2$ on the SFNs with $\alpha = 5.5$ and $k_m = 4$, supplementing Fig.~\ref{fig:fig2}.

\section{Comparison between theory and numerics}

\begin{figure}[h]
\centering
\includegraphics[width=0.36\textwidth]{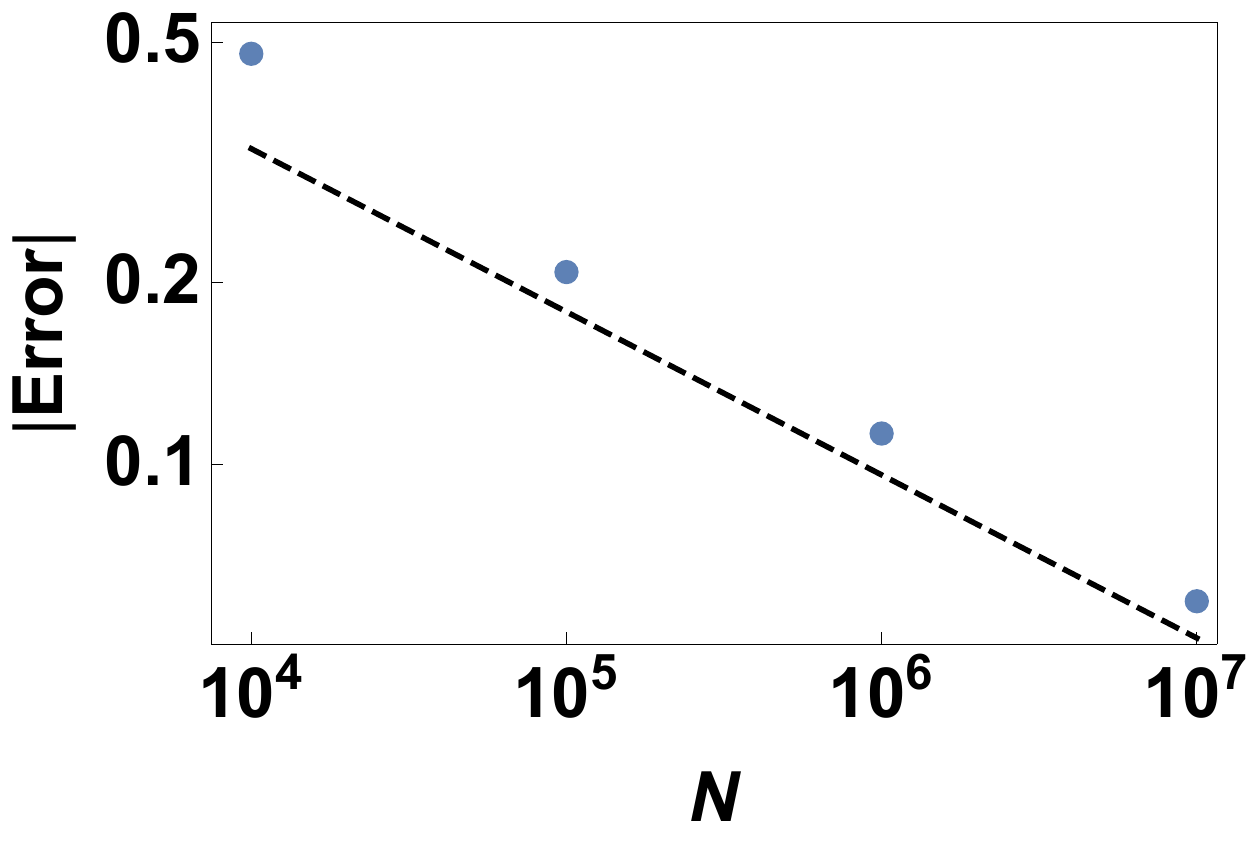}~
\includegraphics[width=0.6\textwidth]{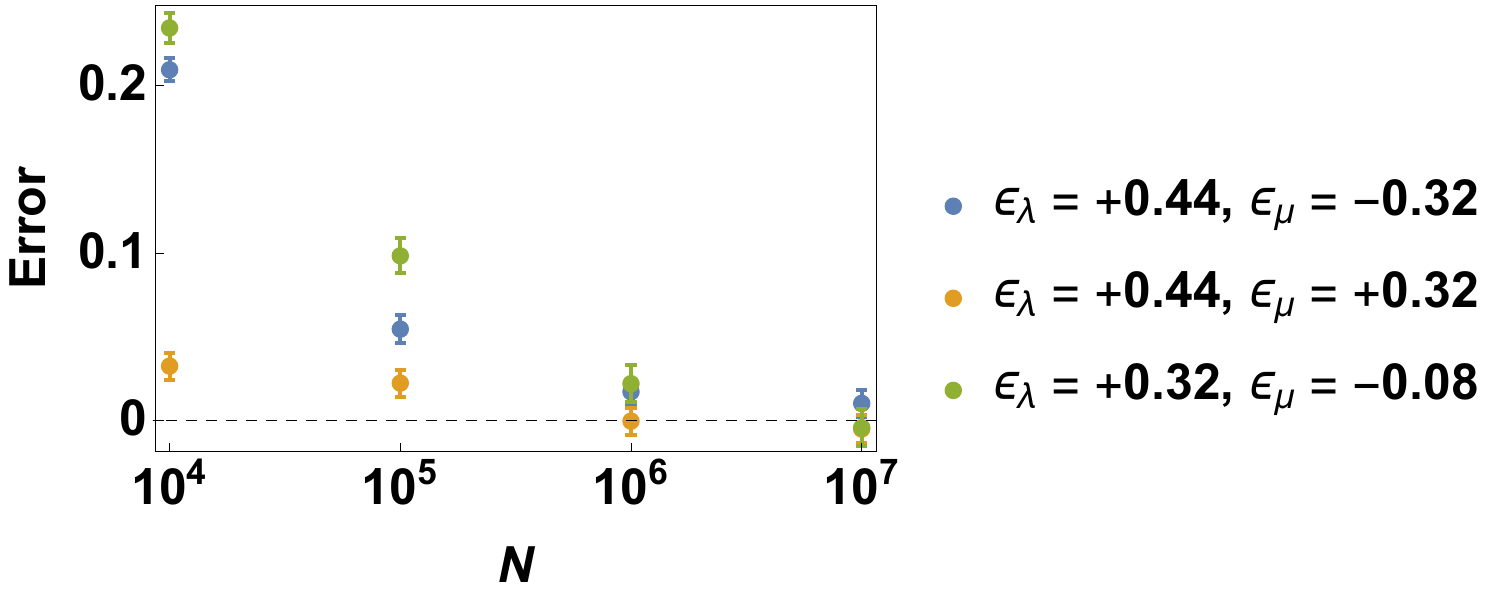}
\caption{\label{fig:figS3} (Left) Error ratio of $\langle R \rangle$ (i.e. $\frac{\text{numerics}}{\text{prediction}}-1$) for scale-free networks with $\alpha = 2.5$ and $k_m = 4$ at $\lambda = 0.11$ and $\mu = 0.08$. The dashed line indicates a power-law decay $N^{-0.27}$. (Right) Error ratio of $\langle R \rangle$ for scale-free networks with $\alpha = 3.5$ and $k_m = 4$. The error bars indicate the range of sampling error.}	
\end{figure}

In Fig.~\ref{fig:figS3}, we show that deviations of the numerical data from the theoretical predictions of $\langle R \rangle$ converge to zero as the network size $N$ increases to infinity.

\end{spacing}

\bibliography{ref-GEP-PRL}

\end{document}